\pdfoutput=1
\documentclass[fleqn,10pt]{SelfArx} 

\usepackage[english]{babel} 

\usepackage[super,square,sort&compress,numbers]{natbib}
\usepackage{url}
\usepackage{setspace} 

\usepackage{etoolbox}

\usepackage{graphicx}

\usepackage{listings}

\usepackage{url}

\usepackage{hyperref}

\usepackage[htt]{hyphenat}

\usepackage{graphicx}

\usepackage{algpseudocode}

\lstdefinestyle{mystyle}{basicstyle=\ttfamily\footnotesize}
\AtBeginEnvironment{quote}{\singlespace\vspace{-\topsep}\small}
\AtEndEnvironment{quote}{\vspace{-\topsep}\endsinglespace}

\graphicspath{ {figures/} }

\definecolor{color1}{RGB}{0,0,90} 
\definecolor{color2}{RGB}{0,20,20} 

\usepackage{hyperref} 
\hypersetup{hidelinks,colorlinks,breaklinks=true,urlcolor=color2,citecolor=color1,linkcolor=color1,bookmarksopen=false,pdftitle={Title},pdfauthor={Author}}

\PaperTitle{JAMPI: efficient matrix multiplication in Spark using Barrier Execution Mode} 

\Authors{Tamas Foldi\textsuperscript{1}*, Chris von Csefalvay\textsuperscript{1}, Nicolas A Perez\textsuperscript{2}} 
\affiliation{\textsuperscript{1}\textit{Starschema Inc., Arlington, VA.}} 
\affiliation{\textsuperscript{2}\textit{Hewlett Packard Enterprise Co.,  San Jose, CA.}} 
\affiliation{*\textbf{Corresponding author}: tfoldi@starschema.net} 

\Abstract{The new barrier mode in Apache Spark allows embedding distributed deep learning training as a Spark stage to simplify the distributed training workflow. In Spark, a task in a stage doesn’t depend on any other tasks in the same stage, and hence it can be scheduled independently. However, several algorithms require more sophisticated inter-task communications, similar to the MPI paradigm. By combining distributed message passing (using asynchronous network IO), OpenJDK's new auto-vectorization and Spark's barrier execution mode, we can add non-map/reduce based algorithms, such as Cannon's distributed matrix multiplication to Spark. We document an efficient distributed matrix multiplication using Cannon's algorithm, which improves significantly on the performance of the existing MLlib implementation. Used within a barrier task, the algorithm described herein results in an up to 24\% performance increase on a 10,000x10,000 square matrix with a significantly lower memory footprint. Applications of efficient matrix multiplication include, among others, accelerating the training and implementation of deep convolutional neural network based workloads, and thus such efficient algorithms can play a ground-breaking role in faster, more efficient execution of even the most complicated machine learning tasks.}

\begin{document}
\maketitle 


\section{Introduction} 
\label{sec:introduction}

The recent decade has seen the emergence of two immensely powerful processes in tandem: the rise of big data handling solutions like Apache Spark on one hand and the apotheosis of deep learning as the tool of choice for demanding computational solutions for machine learning problems. Yet at its essence, big data and deep learning remain not only separate communities but also significantly separate domains of software. Despite deep learning over big data becoming a crucial tool in a range of applications, including in computer vision,\cite{guo2016deep,voulodimos2018deep}  bioinformatics,\cite{spencer2014deep,alipanahi2015predicting,zhang2016deep,wei2018prediction} natural language processing (NLP),\cite{deselaers2009deep,socher2012deep,young2018recent,otter2020survey} clinical medicine,\cite{bar2015chest,havaei2016deep,liu2017detecting,stead2018clinical,campanella2019clinical,lehman2019mammographic} anomaly detection in cybersecurity and fraud detection,\cite{du2017deeplog,shone2018deep,chalapathy2019deep} and collaborative intelligence/recommender systems,\cite{wang2015collaborative,deng2016deep,karatzoglou2017deep,batmaz2019review} its full potential remains to be harnessed. The primary impediment in this respect is largely a divergence of attitudes and concerns, leading to two divergent paradigms of development:

\begin{itemize}
	\item \emph{The big data paradigm}, primarily designed around RDDs and the the DataFrame-based API. This outlook has dominated the development of Apache Spark.
	\item \emph{The DL/ML paradigm}, which is primarily focused on efficient linear algebra operations to facilitate machine learning approaches, especially matrix algebra for deep neural networks.
\end{itemize}

The future of deep learning over big data depends greatly on facilitating the convergence of these two worlds into a single, unified paradigm: the use of well-designed big data management tools, such as Apache Spark, to interoperate with the demands of deep learning. The road towards this convergence depends on the development of efficient matrix primitives that facilitate rapid calculations over distributed networks and large data sets.

\begin{figure}
	\centering
	\includegraphics[width=0.9\linewidth]{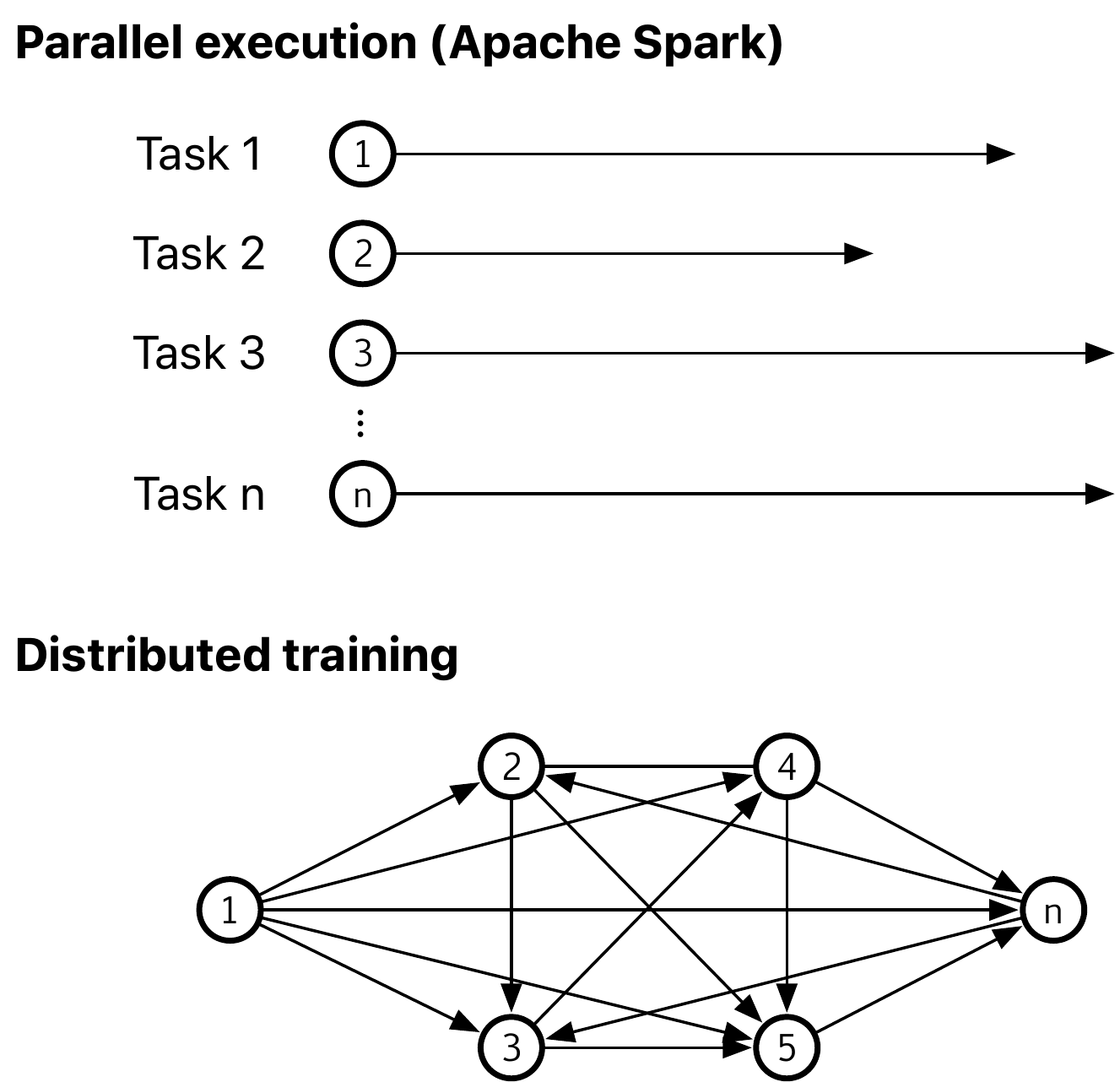}
	\vspace{14pt}
	\caption{Comparative execution models: Apache Spark versus distributed training for neural networks.}
	\label{fig:execution-models}
\end{figure}

The current execution model of Apache Spark is principally focused on independent, embarrassingly parallel tasks that are run and scaled, but the needs of deep learning are primarily focused on distributed training: the performance of completely communicating and coordinating tasks, optimized for interconnectivity rather than independent parallel running, while also maintaining scalability and efficiency. With the recent introduction of the barrier execution mode in Apache Spark, it has finally become possible to construct a computational approach that allows for such networked execution to take place, facilitating distributed training of deep neural networks (see Figure~\ref{fig:execution-models}).

JAMPI (Java Assisted Matrix Product with Inter-task communication), the framework described in this paper, is an efficient and rapid solution to an aspect of efficient matrix primitives, namely matrix multiplication. By integrating JDK's new \texttt{Vector API}, asynchronous network IO (\texttt{nio}) for distributed message passing and Spark's barrier mode, a pure Scala implementation of Cannon's 2.5D matrix multiplication algorithm can be devised that is significantly more efficient than \texttt{MLlib}'s \texttt{BlockMatrix.multiply} function. JAMPI thus avoids reliance on foreign, low level or native code in combination with \texttt{JNI} on one hand, being a pure Scala implementation. On the other hand, it provides a pre-written framework that integrates with Spark as a native task rather than an external MPI procedure call, and handles inter-task communication directly, yielding performance benefits that would otherwise be associated with a low-level MPI implemented resource negotiation framework.

\subsection{Cannon's algorithm} 
\label{sub:cannon_s_algorithm}

Matrix multiplication plays a significant role in a range of practical applications, including (but not limited to) scientific computing, non-linear modeling, agent-based models and the training of deep convolutional neural networks (deep learning). The proliferation of deep learning as the cognitive technology of choice for problems with large source data sets and high-dimensional or high-order multivariate data means that efficiency gains in the underlying linear algebra primitives has the potential to enable significant performance benefits in a wide range of use cases. In particular, constructing primitives that leverage computational capacity through rapid parallel computation and efficient interchange lends itself as an avenue towards these performance gains. While packages comprising efficient matrix primitives already exist,\cite{chetlur2014cudnn} these often operate at a low level and do not integrate well with existing and proven solutions to manage large computational loads.

The matrix multiplication operation $\star$ for an $p \times q$ matrix $\mathbf{A}$ and an $q \times r$ matrix $B$ is defined so that for the resultant matrix $\mathbf{C} = \mathbf{A} \star \mathbf{B}$, each element $c_{i, j}$ is the dot product of the $i$-th row of $\mathbf{A}$ and the $j$-th column of $\mathbf{B}$, i.e.

\begin{equation}
	c_{i, j} = \sum_{k = 1}^n a_{i, k} b_{k, j}
\end{equation}

The multiplication of square matrices constitutes a special case. For a square matrix of order $n$, i.e. an $n \times n$ matrix, a special case obtains, which can be resolved efficiently using Cannon's algorithm.\cite{cannon1969cellular} 

For a square matrix of order $n$, i.e. $n \times n$, Cannon's algorithm uses a toroidally connected mesh $\mathbf{P}^{n \times n}$ of $n^2$ processes. Rendered in pseudocode, the algorithm can be expressed as follows for $p$ processors:

\begin{algorithmic}
	\ForAll{i = 0 : $\sqrt{p}$ - 1}
		\State CShift left A[i; :] by i 
	\EndFor
	
	\ForAll{j = 0 : $\sqrt{p}$ - 1}
		\State CShift up B[:; j] by j 
	\EndFor
	
	\For{k = 0 : $\sqrt{p}$ - 1}
		\For{i = 0 : $\sqrt{p}$ - 1, j = 0 : $\sqrt{p}$ - 1}
			\State C[i, j] += A[i, j] * B[i, j]
			\State CShift left A[i; :] by 1
			\State CShift up B[:; j] by 1
		\EndFor
	\EndFor
\end{algorithmic}

Cannon's algorithm is designed to be performed on a virtual square grid $\mathbf{P}$ of $p$ processors (i.e. a $\sqrt{p} \times \sqrt{p}$ matrix). The multiplicand and multiplier matrices $\mathbf{A}$ and $\mathbf{B}$ are laid out on $\mathbf{P}$, after which the $i$-th row of $\mathbf{A}$ is circularly shifted by $i$ to the left and the $j$-th column of $\mathbf{B}$ circularly shifted by $j$ elements up. Then, $n$ times, the two entries mapped onto $p_{i, j}$ are multiplied and added onto the running value of $p_{i, j}$, after which each row of $\mathbf{A}$ is shifted left by one element and each column of $\mathbf{B}$ is shifted up by one element.

Standard methods of multiplying dense matrices require $O(n^3)$ floating operations for an $n \times n$ matrix. Cannon's algorithm improves on this by reducing it to $O(\frac{n^3}{p})$. In particular, because of the fact that memory is not dependent on the number of processors, it scales dynamically with the number of processors. This makes it an attractive candidate for implementation as a high-performance distributed matrix multiplication primitive.


\subsection{Spark's barrier mode} 
\label{sub:spark_s_barrier_mode}

Spark's barrier mode is a new mode of execution introduced to Apache Spark as part of Project Hydrogen.\cite{projecthydrogensite} Barrier execution features gang scheduling on top of the MapReduce execution model to support distributed deep learning tasks that are executed or embedded as Spark steps. The current implementation ensures that all tasks (limited to \texttt{mapPartitions}) are executed at the same time, and collectively cancels and restarts all tasks in case of failure events. In addition to true parallel execution, the workers' host names and partition identifiers are accessible inside the tasks, alongside a \texttt{barrier} call, similar to MPI's \texttt{MPI\_Barrier} function.\cite{projecthydrogenpres}

While this functionality is sufficient to support the primary use case of Spark's barrier mode -- namely, executing embedded MPI or other foreign, i.e. non-Spark and non-JVM, steps within a Spark application --, it does not provide any inter-task communication primitive to implement the same algorithms within JVM/Spark native steps. In fact, the design documentation for Spark's barrier mode clearly defines this as outside the scope of the project, stating that beyond a simple \texttt{BarrierTaskContext.barrier()} call, no intra-communication functionality will be part of the implementation. It is assumed that such functionality would be handled by the user program. It is our view based on our extensive experience with implementing deep learning solutions on distributed systems that this is a clear show-stopper: if Spark is to be a force to be reckoned with as the data layer for deep learning applications over big data, it should not force execution outside Spark's boundaries.



\section{Methods} 
\label{sec:methods}

\subsection{Cannon's algorithm on MPI} 
\label{sub:cannon_s_algorithm_on_mpi}

The MPI version of the algorithm described in Subsection~\ref{sub:cannon_s_algorithm} relies on MPI's Cartesian topology. After setting up a 2D communication grid of processors with \texttt{MPI\_Cart\_create}, processors exchange data with their neighbors by calling \texttt{MPI\_Sendrecv\_replace}. In the main loop, each processor executes a local dot product calculation, then shifts the results horizontally for matrix \texttt{a} and vertically for matrix \texttt{b}. In our benchmarks, we used \texttt{MPICH} version 3.3.2 as the underlying MPI implementation.

To speed up matrix multiplication, we applied \texttt{-O4} \texttt{-ftree-vectorize} \texttt{-march=native} GNU C compiler flags to ensure vectorized code execution. By vectorization, we refer using SIMD (Single Instruction, Multiple Data) CPU features, more precisely Advanced Vector Extensions (AVX-512F) that allows faster execution of fused multiply–add (\texttt{{FMAC}}) operations in local/partial matrix dot product steps. After compiling our code with GCC 7.3.1 we ensured that the disassembled code contains \texttt{vfmadd231sd} instruction for vectorized \texttt{FMAC}. 


\subsection{JAMPI} 
\label{sub:jampi_implementation}

JAMPI is a \emph{de novo} native Scala implementation of Cannon's algorithm as described in described in Subsection~\ref{sub:cannon_s_algorithm}. For message passing, we built an \texttt{nio} based asynchronous message passing library that mimics MPI's Cartesian topology and send-receive-replace functionality. To avoid unnecessary memory copies and to optimize performance for both throughput and latency, our \texttt{PeerMessage} object allocates fixed 8MB off-heap buffers for both sending and receiving data. Send and receive network operations are executed asynchronously and in parallel.

The matrix multiplication is embedded into a barrier execution task, which is parametrized by the the number of partitions, the local partition ID, the hostnames for the other partitions (\texttt{address} from \texttt{BarrierTaskContext.getTaskInfos()}), as well as the  the local matrix pairs from the RDD.

\begin{lstlisting}
def dotProduct[T : ClassTag](
  partitionId: Integer,
  numOfPartitions: Integer,
  hostMap: Array[String],
  matrixA: Array[T],
  MatrixB: Array[T]): Array[T] 
\end{lstlisting}

JAMPI supports \texttt{double}, \texttt{float} and \texttt{int} Java primitive data types passed as Java \texttt{Array}s.


\subsection{Vectorization using Panama OpenJDK} 
\label{sub:vector_panama}

In order to achieve performance on par with the optimized MPI implementation for local dot product steps, we used JVM's native vector intrinsics and super-word optimization capabilities for both JAMPI and MLlib Spark application benchmarks. The most recent and most comprehensive vectorization support in JVM is found in the \texttt{Vector API} module, part of OpenJDK's \emph{Project Panama}. While the \texttt{Vector API} module is currently in incubation status, we consider it stable enough to use for both the Spark platform and application code.

For fair benchmarking, we avoided using \texttt{Vector<>} objects or advanced methods such as manual unrolling. While these techniques could potentially further improve performance, our goals were to compare the distributed algorithms' performance with the same CPU opcodes used in local matrix multiplications. From the JIT compiler outputs, we confirmed that both Spark applications were using \texttt{vfmadd231sd}, just as in the GCC compiled MPI version.

To use the new vector intrinsics' features, we built a custom OpenJDK package from the tip of the \texttt{panama/dev} branch (\texttt{dev-442a69af7bad}). The applied JVM flags were \texttt{--add-modules jdk.incubator.vector} and \texttt{ -XX:TypeProfileLevel=121} for both JAMPI and MLlib applications. 



\subsection{Apache Spark MLlib}

We used Apache Spark MLlib's built-in \texttt{BlockMatrix.multiply()} as a baseline to compare with JAMPI's speed and resource usage. It is known that MLlib's implementation is often faster if the number of partitions exceeds that of worker cores (typically by a factor of 2-4 at least), a scenario known as \emph{over-partitioning}. To ensure that this is adequately reflected, we performed two test runs – a 'normal' test run, where partitions are set to equal the number of worker cores, and an 'over-partitioned' test run, where partitions equal four times the number of worker cores.

\begin{figure*}
	\centering
	\includegraphics[width=0.9\linewidth]{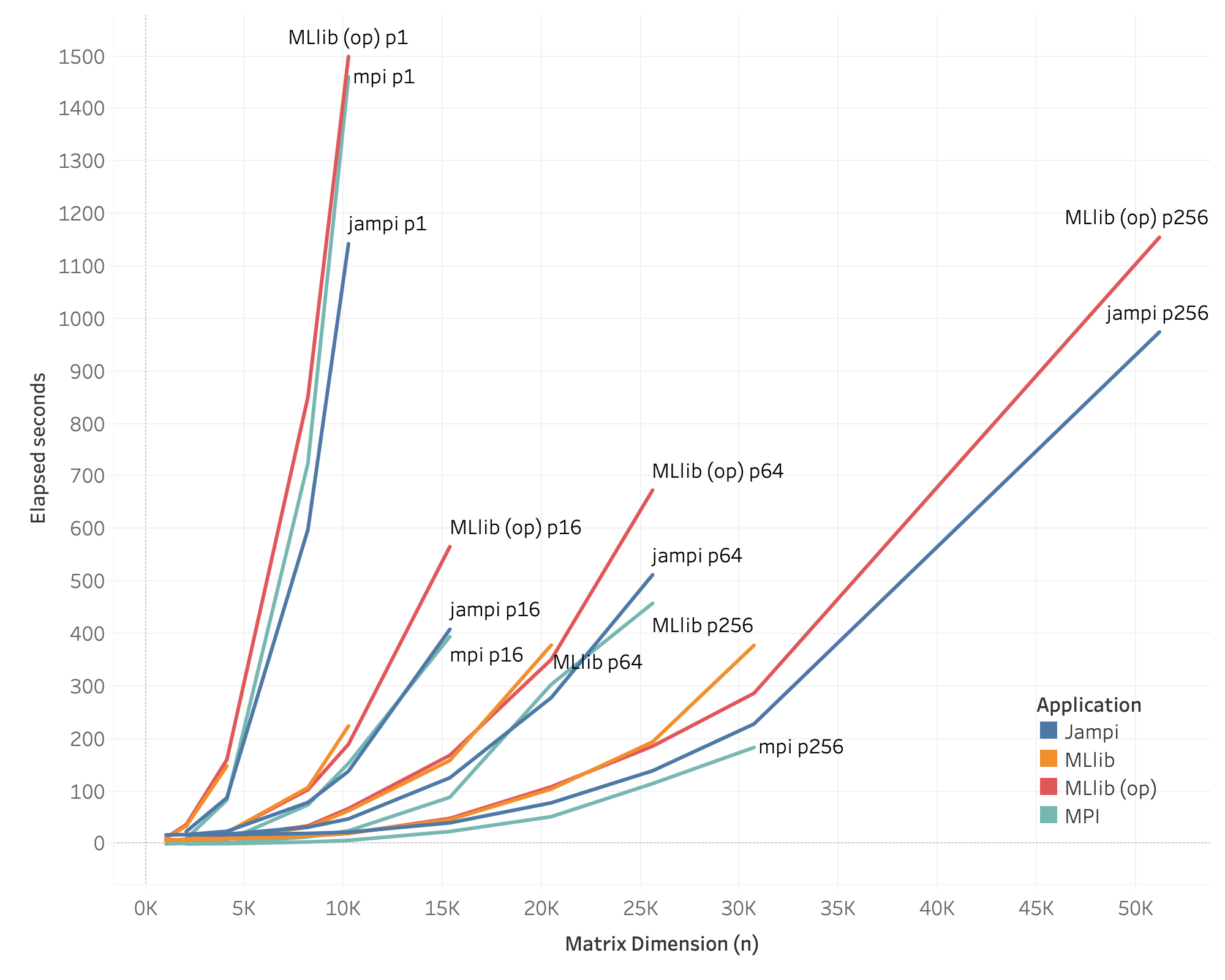}
	\vspace{14pt}
	\caption{Comparative performance of JAMPI, native MPI and MLlib on random matrices of various dimensions, on 1, 16, 64 and 256 cores.}
	\label{fig:overall_performance}
\end{figure*}
\subsection{Test protocols} 
\label{sub:test_protocols}

All tests were performed on Amazon Web Services EC2 instances using \texttt{m5} instance types with Intel\textsuperscript{\textregistered} Xeon\textsuperscript{\textregistered} Platinum 8175M CPUs and 4GB RAM per core. Tests were conducted on Apache Spark 3.0.0-preview2 with a separate master node. The driver process was initiated from the master node, and its resource consumption is not included in the results. For single core tests, 2-core CPUs were used, with the second CPU core having been manually disabled in the VM.

\begin{center}
	\begin{tabular}{ |c|c|c|c| } 
	 \hline
	 Total worker cores & Instance type & Nodes & Partitions  \\ 
	 \hline\hline
	 1 & m5.large & 1 & 1 \\ 
	 \hline
	 16 & m5.xlarge & 4 & 4 \\ 
	 \hline
	 64 & m5.2xlarge & 8 & 8 \\ 
	 \hline
	 256 & m5.2xlarge & 32 & 8 \\ 
	 \hline
	\end{tabular}
\end{center}

Applications reported only the dot product execution time. A single one-value reducer (\texttt{avg}) was included to trigger RDD reduce/collection on Spark without moving substantial amount of data to the driver process. Timings thus exclude the MPI and Spark application startup times, but included the time required to establish a barrier task step during the RDD reduce step. For testing, random matrices composed of 64-bit floating point elements were used. Test scenarios were performed ten times, capturing execution time, CPU and memory consumption. Test scenarios, as well as the original JAMPI source code, are available online on Github under \texttt{https://github.com/starschema/jampi-spark-dotmatrix}.


\subsection{Scalability analysis} 
\label{sub:scalability_analysis}

An important aspect of any distributed algorithm is its ability to scale up as the problem size increases. This is crucial for proving the value of an algorithmic solution, since it demonstrates its ability to solve increasingly complex instances of the same fundamental problem effectively. There are intrinsic issues when scaling distributed multi-processor algorithms. It is known, for instance, that the memory requirement for each processor increases as we add processors to a computation. Therefore, we must analyze the effect of problem size on the memory requirements per processor. 

For Cannon's algorithm multiplying two square matrices of size $n \times n$, the problem size $W$ is on the order of $n^2$, i.e.,

\begin{equation}
	W = \mathcal{O}(n^2)
	\label{eq:problem_size}
\end{equation}

The sequential time, that is when $p$ = 1, is 

\begin{equation}
	T_1(n) = \mathcal{O}(n^3)
	\label{eq:seq_runtime}
\end{equation}

For $p$ processors, the execution time for a matrix of size $n \times n$ is given as $T_p (n)$. It follows that parallelization of the problem yields a speed-up calculated as $\frac{W}{T_p (n)}$. 

In addition, parallel execution of an $n \times n$ problem size over $p$ processors will incur a performance overhead of $T_o(n, p)$, including all communication costs. 

It is known that the communication cost $D$, which is how much data is being shifted across the $p$ processors, can be calculated as 

\begin{equation}
	D = O(\frac{n^2}{\sqrt{p}})
\end{equation}

\noindent Using the following iso-efficiency relationship of parallel systems,  

\begin{equation}
	T_1(n) \geq c \ T_o(n, p)
	\label{eq:iso_efficiency}
\end{equation}

\noindent Substituting Equation~\eqref{eq:seq_runtime} in Equation~\eqref{eq:iso_efficiency}, it follows that

\begin{equation}
	n^3 \geq c \ \sqrt{p} \ n^2 \ \implies \ n \geq c \ \sqrt{p}
	\label{eq:scale_factor}
\end{equation}

\noindent It follows thus from Equation~\eqref{eq:scale_factor} and the definition of $W$ in Equation~\eqref{eq:problem_size} that

\begin{equation}
	\frac{M(c \ \sqrt{p})}{p} = \frac{c^2 \ p}{p} = c^2
\end{equation}

More generally, it holds that for a problem size $W$ and $p$ processors, Cannon's Algorithm memory requirements increase by a constant factor $c^2$ that is independent of the number of processors $p$ involved in the computation.  Since the memory requirements per processor increase linearly, without direct relationship to $p$, it can be said that Cannon's algorithm is extremely scalable. 

\begin{figure}
	\centering
	\includegraphics[width=0.9\linewidth]{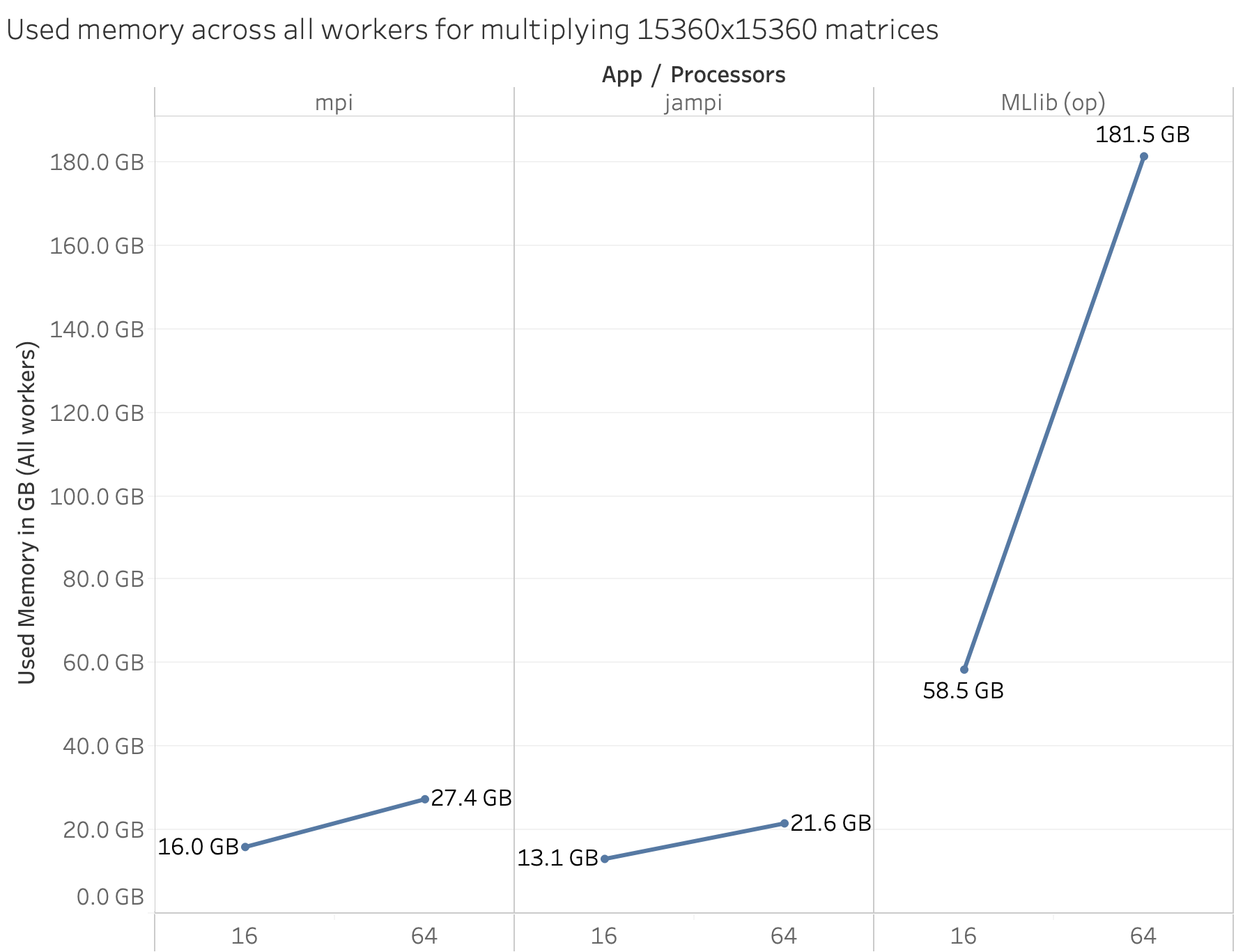}
	\vspace{14pt}
	\caption{Comparative memory usage between JAMPI, MPI and MLlib}
	\label{fig:memory_usage}
\end{figure}

Figure~\ref{fig:memory_usage} illustrates this scaling behavior comparatively between JAMPI, a pure MPI implementation and MLlib. JAMPI, as well as the MPI algorithm test case, are both direct implementations of Cannon's algorithm, thus having the same scalability behavior. 

It is evident from Figure~\ref{fig:memory_usage} that MLLib's memory requirement increases quite fast, suggesting that its scalability factor is larger than that of Cannon's algorithm (i.e. it is less scalable). This is a key limitation of MLlib and Spark when compared to MPI and JAMPI alike, which scale better. Indeed, in some test scenarios, we have been unable to scale MLlib beyond a certain problem size, indicating that in addition to its poor performance compared to MPI and JAMPI, it is also limited in the maximum problem size it can accommodate with a set level of resources. Neither JAMPI nor the native MPI implementation is so limited.


\section{Results} 
\label{sec:results}

Comparative analysis of runtimes over a range of matrix sizes reveals that JAMPI is significantly superior to MLlib, even when over-partitioned (see Figure~\ref{fig:comparison_to_jampi}, over-partitioning is denoted by \texttt{op}). When normalized against JAMPI's execution times over 16 and 64 cores, execution time is slower for smaller matrices (under 4096 $\times$ 4096 elements) due to the need to establish and run the barrier execution task. However, beyond a trivial problem size, JAMPI and the MPI implementation rapidly become significantly more efficient, regardless of the number of cores. Notably, plain MLlib (i.e. without over-partitioning) was unable to accommodate a problem size beyond 10240 $\times$ 10240 (for 16 cores) or 20480 $\times$ 20480 (for 64 cores).

\subsection{Memory usage} 
\label{sub:memory_usage}

Memory usage has been a documented limiting factor, with pure MLlib reaching execution limits at relatively trivial matrix dimensions per processor (Table~\ref{tab:out_of_memory}). While over-partitioning slightly increases the maximum matrix size, MLlib suffers from not only lower performance but also a memory consumption upper bound that limits its ability to scale to larger problem sizes.

Our research indicates that for a 10,240 $\times$ 10,240 element standard matrix, JAMPI and MPI perform approximately equally (4,889 MB vs 5,108 MB, respectively, for 256 cores), while both over-partitioned and regular MLlib execution creates a marginally larger memory footprint (6,049 MB and 6,423 MB, respectively, for 256 cores).  However, with increasing problem size, differences become vastly apparent: for a 30,720 $\times$ 30,720 element matrix, MPI and JAMPI continue to require a constant memory footprint (5,572 MB and 6,084 MB, respectively), while the same problem size requires 24,525 MB with over-partitioning and 29,445 MB without. In other words, JAMPI and MPI memory burden increases constantly regardless of the number of cores, while MLLib's memory consumption increases rapidly, as Figure~\ref{fig:memory_usage} indicates. For instance, when processing 30,720 $\times$ 30,720 matrix size, MLlib requires a 4.03 (with over-partitioning) to 4.84 (without over-partitioning) times larger memory allocation.

Comparative analysis of memory usage (see Figure~\ref{fig:memory_usage}) shows that JAMPI is generally on par (within 30\%) of the pure MPI implementation, while MLlib typically requires approximately four times the amount of memory allocation that the MPI based approaches demand, with regular MLlib requiring typically 15\% to 50\% more memory than over-partitioned implementations.

\begin{table}
	\centering
	\begin{tabular}{ |c|c|c| } 
	 \hline
	 Cores & MLlib & MLlib (op)  \\ 
	 \hline\hline
	 1 & 4096 & 10240 \\ 
	 \hline
	 16 & 10240 & 15360 \\ 
	 \hline
	 64 & 20480 & 25600 \\ 
	 \hline
	 256 & 30720 & 51200 \\ 
	 \hline
	\end{tabular}
	\caption{Out-of-memory boundary sizes for MLlib, in normal (MLlib) and over-partitioned (MLlib (op)) mode.}
	\label{tab:out_of_memory}
\end{table}
	

\subsection{Performance} 
\label{sub:performance}

\begin{figure}
	\centering
	\includegraphics[width=0.9\linewidth]{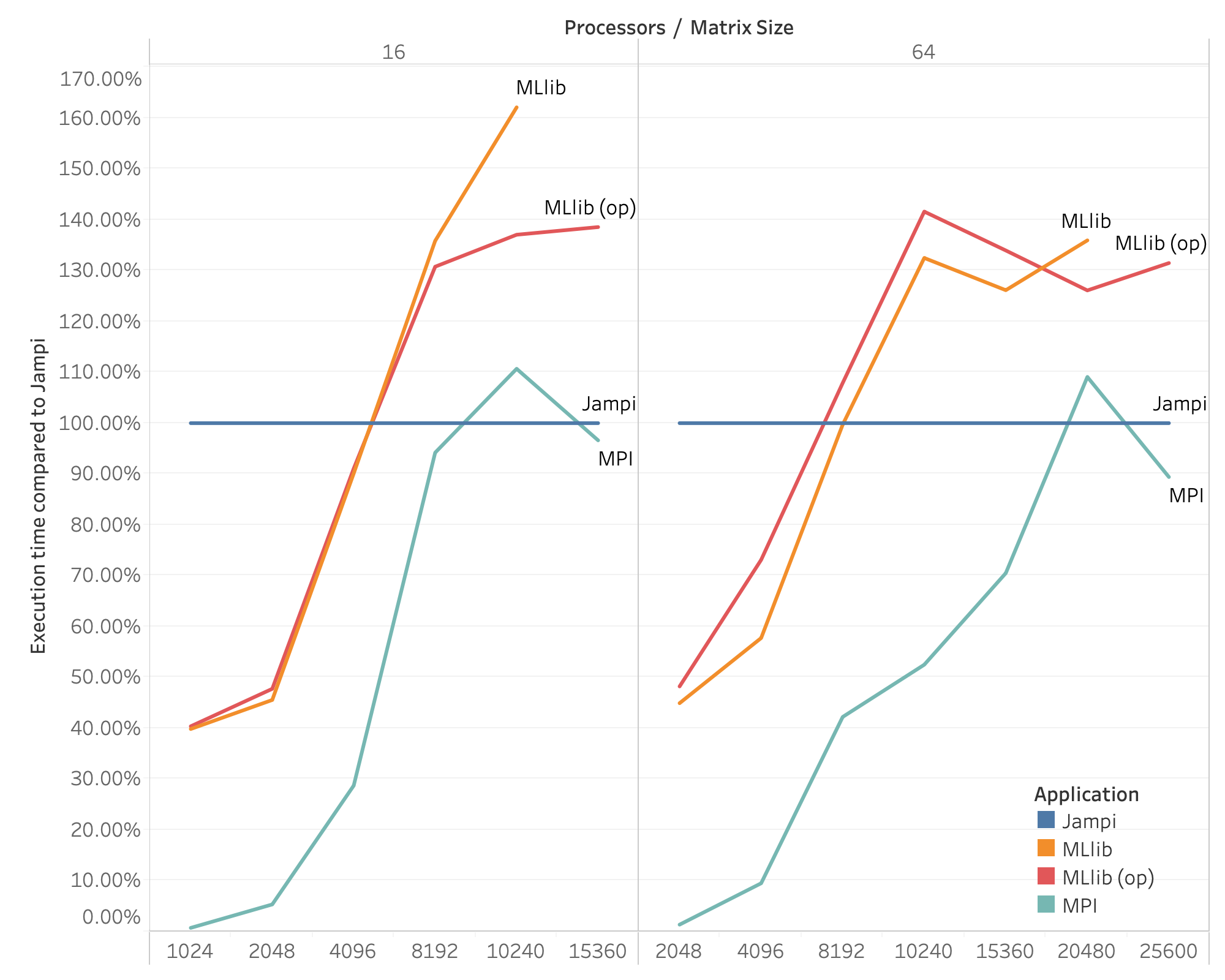}
	\vspace{14pt}
	\caption{Comparative execution times on 16 and 64 cores for various matrix sizes, normalized to JAMPI (blue).}
	\label{fig:comparison_to_jampi}
\end{figure}


Comparing performance in terms of execution time shows a similar picture in all multi-core environments. MLlib, both with and without over-partitioning, presents a lower execution time compared to JAMPI in trivial-sized matrices (4096 $\times$ 4096 for 16- and 64-core environments, 10240 $\times$ 10240 for 256-core environments). 

However, MLlib execution times rapidly increases, yielding an average of 29.52\% (with over-partitioning) to 54.54\% (without over-partitioning) slower execution time at the largest feasible matrix for the given number of cores. On the other hand, for large matrices, as Figure~\ref{fig:comparison_to_jampi} shows, a pure MPI implementation is typically 3.38\% to 19.59\% faster than JAMPI. 

The comparative analysis of performance indicators shows that while a pure MPI implementation is somewhat faster than JAMPI, this difference is significantly smaller than the difference between the MLlib implementation and JAMPI, proving that JAMPI is an efficient and fast alternative to pure MPI applications without a significant performance overhead.



\section{Conclusion} 
\label{sec:conclusion}

Cannon's algorithm can be implemented quite conveniently using a barrier task within Spark, providing a native interpretation of this highly efficient distributed linear algebra primitive. By using barrier tasks to reimplement matrix primitives with Panama's built-in efficient vectorization and asynchronous communication (as provided by \texttt{nio} in this case), very significant performance gains can be effected on frequently used tasks. The proposed implementation of Cannon's algorithm, for instance, has yielded an almost 25\% decrease in execution time, and has been superior to the MLlib implementation on all core sizes above trivial matrix sizes. While this algorithm is limited to square matrices, the general effectiveness gains are indicative of a strong theoretical and practical benefit of further research in ways efficient matrix primitives can be integrated with big data solutions like Apache Spark. Further research in this field is required to create a coherent stack of matrix primitives in order to allow modern deep learning applications, relying greatly on such building blocks, to leverage the performance benefits of big data solutions in storing and managing data as a layer of an integrated framework of large-scale machine learning.





\phantomsection
\section*{Competing interests} 


The authors have declared no competing interests.

\phantomsection
\section*{Funding statement} 


The research summarized in this paper was funded by Starschema Inc.

\phantomsection

\bibliography{bibliography}

\begin{thebibliography}{27}
\providecommand{\natexlab}[1]{#1}
\providecommand{\url}[1]{\texttt{#1}}
\expandafter\ifx\csname urlstyle\endcsname\relax
  \providecommand{\doi}[1]{doi: #1}\else
  \providecommand{\doi}{doi: \begingroup \urlstyle{rm}\Url}\fi

\bibitem[Guo et~al.(2016)Guo, Liu, Oerlemans, Lao, Wu, and Lew]{guo2016deep}
Yanming Guo, Yu~Liu, Ard Oerlemans, Songyang Lao, Song Wu, and Michael~S Lew.
\newblock Deep learning for visual understanding: A review.
\newblock \emph{Neurocomputing}, 187:\penalty0 27--48, 2016.

\bibitem[Voulodimos et~al.(2018)Voulodimos, Doulamis, Doulamis, and
  Protopapadakis]{voulodimos2018deep}
Athanasios Voulodimos, Nikolaos Doulamis, Anastasios Doulamis, and Eftychios
  Protopapadakis.
\newblock Deep learning for computer vision: A brief review.
\newblock \emph{Computational {I}ntelligence and {N}euroscience}, 2018, 2018.

\bibitem[Spencer et~al.(2014)Spencer, Eickholt, and Cheng]{spencer2014deep}
Matt Spencer, Jesse Eickholt, and Jianlin Cheng.
\newblock A deep learning network approach to ab initio protein secondary
  structure prediction.
\newblock \emph{IEEE/ACM {T}ransactions on {Computational} {Biology} and
  {Bioinformatics}}, 12\penalty0 (1):\penalty0 103--112, 2014.

\bibitem[Alipanahi et~al.(2015)Alipanahi, Delong, Weirauch, and
  Frey]{alipanahi2015predicting}
Babak Alipanahi, Andrew Delong, Matthew~T Weirauch, and Brendan~J Frey.
\newblock Predicting the sequence specificities of {D}{N}{A}-and
  {R}{N}{A}-binding proteins by deep learning.
\newblock \emph{Nature {biotechnology}}, 33\penalty0 (8):\penalty0 831--838,
  2015.

\bibitem[Zhang et~al.(2016)Zhang, Zhou, Hu, Gong, Chen, Cheng, and
  Zeng]{zhang2016deep}
Sai Zhang, Jingtian Zhou, Hailin Hu, Haipeng Gong, Ligong Chen, Chao Cheng, and
  Jianyang Zeng.
\newblock A deep learning framework for modeling structural features of
  {R}{N}{A}-binding protein targets.
\newblock \emph{{N}ucleic {A}cids {R}esearch}, 44\penalty0 (4):\penalty0
  e32--e32, 2016.

\bibitem[Wei et~al.(2018)Wei, Ding, Su, Tang, and Zou]{wei2018prediction}
Leyi Wei, Yijie Ding, Ran Su, Jijun Tang, and Quan Zou.
\newblock Prediction of human protein subcellular localization using deep
  learning.
\newblock \emph{Journal of Parallel and Distributed Computing}, 117:\penalty0
  212--217, 2018.

\bibitem[Deselaers et~al.(2009)Deselaers, Hasan, Bender, and
  Ney]{deselaers2009deep}
Thomas Deselaers, Sa{\v{s}}a Hasan, Oliver Bender, and Hermann Ney.
\newblock A deep learning approach to machine transliteration.
\newblock In \emph{Proceedings of the {Fourth} {Workshop} on {Statistical}
  {Machine} {Translation}}, pages 233--241. Association for Computational
  Linguistics, 2009.

\bibitem[Socher et~al.(2012)Socher, Bengio, and Manning]{socher2012deep}
Richard Socher, Yoshua Bengio, and Chris Manning.
\newblock Deep learning for {NLP}.
\newblock \emph{{T}utorial at {A}ssociation of {C}omputational {L}ogistics
  ({A}{C}{L})}, 2012.

\bibitem[Young et~al.(2018)Young, Hazarika, Poria, and
  Cambria]{young2018recent}
Tom Young, Devamanyu Hazarika, Soujanya Poria, and Erik Cambria.
\newblock Recent trends in deep learning based {Natural} {Language}
  {Processing}.
\newblock \emph{IEEE {Computational} {Intelligence} {Magazine}}, 13\penalty0
  (3):\penalty0 55--75, 2018.

\bibitem[Otter et~al.(2020)Otter, Medina, and Kalita]{otter2020survey}
Daniel~W Otter, Julian~R Medina, and Jugal~K Kalita.
\newblock A survey of the usages of deep learning for natural language
  processing.
\newblock \emph{IEEE {Transactions} on {Neural Networks} and {Learning
  Systems}}, 2020.

\bibitem[Bar et~al.(2015)Bar, Diamant, Wolf, Lieberman, Konen, and
  Greenspan]{bar2015chest}
Yaniv Bar, Idit Diamant, Lior Wolf, Sivan Lieberman, Eli Konen, and Hayit
  Greenspan.
\newblock Chest pathology detection using deep learning with non-medical
  training.
\newblock In \emph{2015 IEEE 12th {International} {Symposium} on {Biomedical}
  {Imaging} (ISBI)}, pages 294--297. IEEE, 2015.

\bibitem[Havaei et~al.(2016)Havaei, Guizard, Larochelle, and
  Jodoin]{havaei2016deep}
Mohammad Havaei, Nicolas Guizard, Hugo Larochelle, and Pierre-Marc Jodoin.
\newblock Deep learning trends for focal brain pathology segmentation in
  {M}{R}{I}.
\newblock In \emph{{Machine} {L}earning for {health} {informatics}}, pages
  125--148. Springer, 2016.

\bibitem[Liu et~al.(2017)Liu, Gadepalli, Norouzi, Dahl, Kohlberger, Boyko,
  Venugopalan, Timofeev, Nelson, Corrado, et~al.]{liu2017detecting}
Yun Liu, Krishna Gadepalli, Mohammad Norouzi, George~E Dahl, Timo Kohlberger,
  Aleksey Boyko, Subhashini Venugopalan, Aleksei Timofeev, Philip~Q Nelson,
  Greg~S Corrado, et~al.
\newblock Detecting cancer metastases on gigapixel pathology images.
\newblock \emph{arXiv preprint arXiv:1703.02442}, 2017.

\bibitem[Stead(2018)]{stead2018clinical}
William~W Stead.
\newblock Clinical implications and challenges of artificial intelligence and
  deep learning.
\newblock \emph{{J}{A}{M}{A}}, 320\penalty0 (11):\penalty0 1107--1108, 2018.

\bibitem[Campanella et~al.(2019)Campanella, Hanna, Geneslaw, Miraflor, Silva,
  Busam, Brogi, Reuter, Klimstra, and Fuchs]{campanella2019clinical}
Gabriele Campanella, Matthew~G Hanna, Luke Geneslaw, Allen Miraflor, Vitor
  Werneck~Krauss Silva, Klaus~J Busam, Edi Brogi, Victor~E Reuter, David~S
  Klimstra, and Thomas~J Fuchs.
\newblock Clinical-grade computational pathology using weakly supervised deep
  learning on whole slide images.
\newblock \emph{Nature {medicine}}, 25\penalty0 (8):\penalty0 1301--1309, 2019.

\bibitem[Lehman et~al.(2019)Lehman, Yala, Schuster, Dontchos, Bahl, Swanson,
  and Barzilay]{lehman2019mammographic}
Constance~D Lehman, Adam Yala, Tal Schuster, Brian Dontchos, Manisha Bahl, Kyle
  Swanson, and Regina Barzilay.
\newblock Mammographic breast density assessment using deep learning: clinical
  implementation.
\newblock \emph{Radiology}, 290\penalty0 (1):\penalty0 52--58, 2019.

\bibitem[Du et~al.(2017)Du, Li, Zheng, and Srikumar]{du2017deeplog}
Min Du, Feifei Li, Guineng Zheng, and Vivek Srikumar.
\newblock Deeplog: Anomaly detection and diagnosis from system logs through
  deep learning.
\newblock In \emph{Proceedings of the 2017 ACM SIGSAC {Conference} on
  {Computer} and {Communications} {Security}}, pages 1285--1298, 2017.

\bibitem[Shone et~al.(2018)Shone, Ngoc, Phai, and Shi]{shone2018deep}
Nathan Shone, Tran~Nguyen Ngoc, Vu~Dinh Phai, and Qi~Shi.
\newblock A deep learning approach to network intrusion detection.
\newblock \emph{IEEE {Transactions} on {Emerging Topics} in {Computational
  Intelligence}}, 2\penalty0 (1):\penalty0 41--50, 2018.

\bibitem[Chalapathy and Chawla(2019)]{chalapathy2019deep}
Raghavendra Chalapathy and Sanjay Chawla.
\newblock Deep learning for anomaly detection: A survey.
\newblock \emph{arXiv preprint arXiv:1901.03407}, 2019.

\bibitem[Wang et~al.(2015)Wang, Wang, and Yeung]{wang2015collaborative}
Hao Wang, Naiyan Wang, and Dit-Yan Yeung.
\newblock Collaborative deep learning for recommender systems.
\newblock In \emph{Proceedings of the 21th ACM SIGKDD {International}
  {Conference} on {Knowledge} {Discovery} and {Data} {Mining}}, pages
  1235--1244, 2015.

\bibitem[Deng et~al.(2016)Deng, Huang, Xu, Wu, and Wu]{deng2016deep}
Shuiguang Deng, Longtao Huang, Guandong Xu, Xindong Wu, and Zhaohui Wu.
\newblock On deep learning for trust-aware recommendations in social networks.
\newblock \emph{IEEE {Transactions} on {Neural} {Networks} and {Learning}
  {Systems}}, 28\penalty0 (5):\penalty0 1164--1177, 2016.

\bibitem[Karatzoglou and Hidasi(2017)]{karatzoglou2017deep}
Alexandros Karatzoglou and Bal{\'a}zs Hidasi.
\newblock Deep learning for recommender systems.
\newblock In \emph{Proceedings of the {E}leventh ACM {conference} on
  {recommender systems}}, pages 396--397, 2017.

\bibitem[Batmaz et~al.(2019)Batmaz, Yurekli, Bilge, and
  Kaleli]{batmaz2019review}
Zeynep Batmaz, Ali Yurekli, Alper Bilge, and Cihan Kaleli.
\newblock A review on deep learning for recommender systems: challenges and
  remedies.
\newblock \emph{{Artificial} {Intelligence} {Review}}, 52\penalty0
  (1):\penalty0 1--37, 2019.

\bibitem[Chetlur et~al.(2014)Chetlur, Woolley, Vandermersch, Cohen, Tran,
  Catanzaro, and Shelhamer]{chetlur2014cudnn}
Sharan Chetlur, Cliff Woolley, Philippe Vandermersch, Jonathan Cohen, John
  Tran, Bryan Catanzaro, and Evan Shelhamer.
\newblock cudnn: Efficient primitives for deep learning.
\newblock \emph{arXiv preprint arXiv:1410.0759}, 2014.

\bibitem[Cannon(1969)]{cannon1969cellular}
Lynn~Elliot Cannon.
\newblock \emph{A cellular computer to implement the {K}alman filter
  algorithm}.
\newblock PhD thesis, Montana State University-Bozeman, College of Engineering,
  1969.

\bibitem[{Jules Damji}()]{projecthydrogensite}
{Jules Damji}.
\newblock \emph{{Bay Area Apache Spark Meetup Summary} at {Databricks} {H}{Q}}.
\newblock URL
  \url{https://databricks.com/blog/2018/07/25/bay-area-apache-spark-meetup-summary-databricks-hq.html}.

\bibitem[Meng(2018)]{projecthydrogenpres}
Xiangrui Meng.
\newblock \emph{{P}roject {H}ydrogen: {S}tate-of-the-{A}rt {D}eep {L}earning on
  {A}pache {S}park}, 2018.
\newblock URL
  \url{https://www.slideshare.net/databricks/project-hydrogen-stateoftheart-deep-learning-on-apache-spark}.

\end{thebibliography}


\end{document}